\documentclass[10pt,final,journal]{IEEEtran}
\ifCLASSINFOpdf
  \usepackage[pdftex]{graphicx}
\else
\fi
\usepackage{amsmath}
\ifCLASSOPTIONcompsoc
 \usepackage[caption=false,font=normalsize,labelfont=sf,textfont=sf]{subfig}
\else
 \usepackage[caption=false,font=footnotesize]{subfig}
\fi
\usepackage{dblfloatfix}

\usepackage[nomarkers]{endfloat}
\let\MYoriglatexcaption\caption
\renewcommand{\caption}[2][\relax]{\MYoriglatexcaption[#2]{#2}}
\usepackage{authblk}
\usepackage{bm}
\usepackage{dsfont}
\usepackage{tabularx}
\usepackage{algorithm}
\usepackage[noend]{algpseudocode}
\usepackage{multirow}
\usepackage[export]{adjustbox}
\usepackage{xcolor}
\usepackage[T1]{fontenc}

\newcommand\Tstrut{\rule{0pt}{2.6ex}}         
\newcommand\Bstrut{\rule[-0.9ex]{0pt}{0pt}}   

\begin{document}
%
\title{An unsupervised transfer learning algorithm for sleep monitoring}
%
%
%

\author[1]{Xichen She}
\author[2]{Yaya Zhai}
\author[3]{Ricardo Henao}
\author[3]{Christopher W. Woods}
\author[3]{Geoffrey S. Ginsburg}
\author[1]{Peter X.K. Song}
\author[4]{Alfred O. Hero, ~\IEEEmembership{Fellow,~IEEE}}
\affil[1]{Department of Biostatistics, University of Michigan, Ann Arbor, MI 48109}
\affil[2]{Department of Computational Medicine and Bioinformatics, University of Michigan, Ann Arbor, MI 48109}
\affil[3]{Center for Applied Genomics and Precision Medicine, Duke University, Durham, NC 27708}
\affil[4]{Departments of Electrical Engineering and Computer Science, Biomedical Engineering, and Statistics, University of Michigan, Ann Arbor, MI 48109}

\markboth{IEEE TRANSACTIONS ON BIOMEDICAL ENGINEERING}%
{She \MakeLowercase{\textit{et al.}}: An unsupervised transfer learning algorithm for sleep monitoring {  under perturbed environments}}
%



\maketitle


\begin{abstract}
  \textbf{\textit{Objective}}: To develop multisensor-wearable-device sleep monitoring algorithms that are robust to health disruptions affecting sleep patterns.  
  \textbf{\textit{Methods}}: We develop an unsupervised transfer learning algorithm based on a multivariate hidden Markov model and Fisher's linear discriminant analysis, adaptively adjusting to sleep pattern shift by training on dynamics of sleep/wake states.
  The proposed algorithm operates, without requiring \textit{a priori} information about true sleep/wake states, by establishing an initial training set with hidden Markov model and leveraging a taper window mechanism to learn the sleep pattern in an incremental fashion. Our domain-adaptation algorithm is applied to a dataset collected in a human viral challenge study to identify sleep/wake periods of both uninfected and infected participants.
  \textbf{\textit{Results}}: 
  The algorithm successfully detects sleep/wake sessions in subjects whose sleep patterns are disrupted by respiratory infection (H3N2 flu virus). Pre-symptomatic features based on the detected periods are found to be strongly predictive of both infection status (AUC = 0.844) and infection onset time (AUC = 0.885), indicating the effectiveness and usefulness of the algorithm.
  \textbf{\textit{Conclusion}}:  Our method can effectively detect sleep/wake states in the presence of sleep pattern shift. 
  \textbf{\textit{Significance}}: Utilizing integrated multisensor signal processing and adaptive training schemes, our algorithm  is able to capture key sleep patterns in ambulatory monitoring, leading to better automated sleep assessment and prediction.\footnote{This research was partially supported by the Prometheus Program of the Defense Advanced Research Projects Agency (DARPA), grant number N66001-17-2-4014.}
\end{abstract}

\begin{IEEEkeywords}
  Data covariate shift, Fisher's linear discriminant analysis, Sleep monitoring, Human viral challenge study, Wearable. 
\end{IEEEkeywords}

%
\IEEEpeerreviewmaketitle

\section{Introduction}
%
%
%
%
\begin{figure*}[!ht]
	\centering
	\subfloat[]{\includegraphics[width=.8\linewidth]{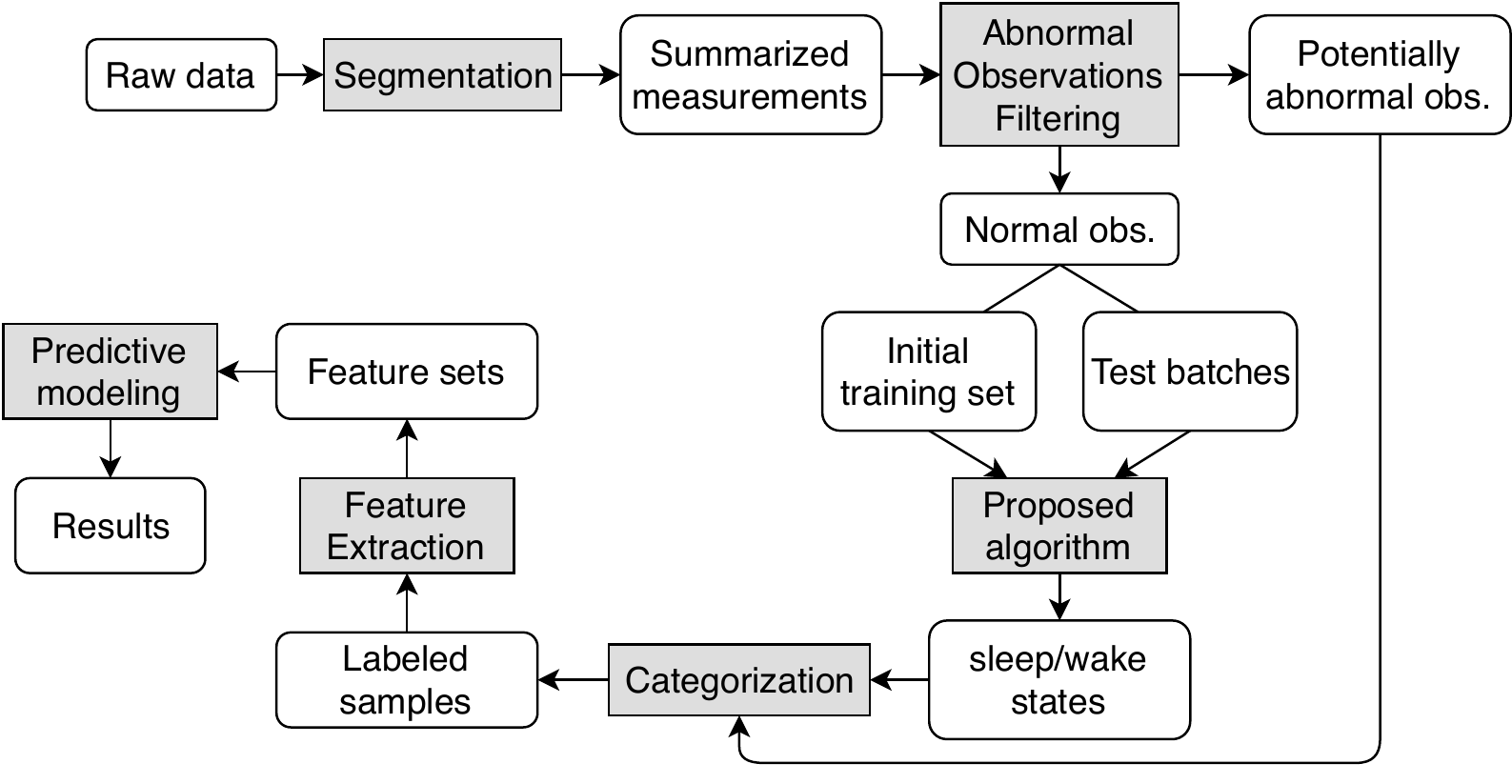}
	\label{fig:pipe}}
	\hfil
	\subfloat[]{\includegraphics[width=.6\linewidth]{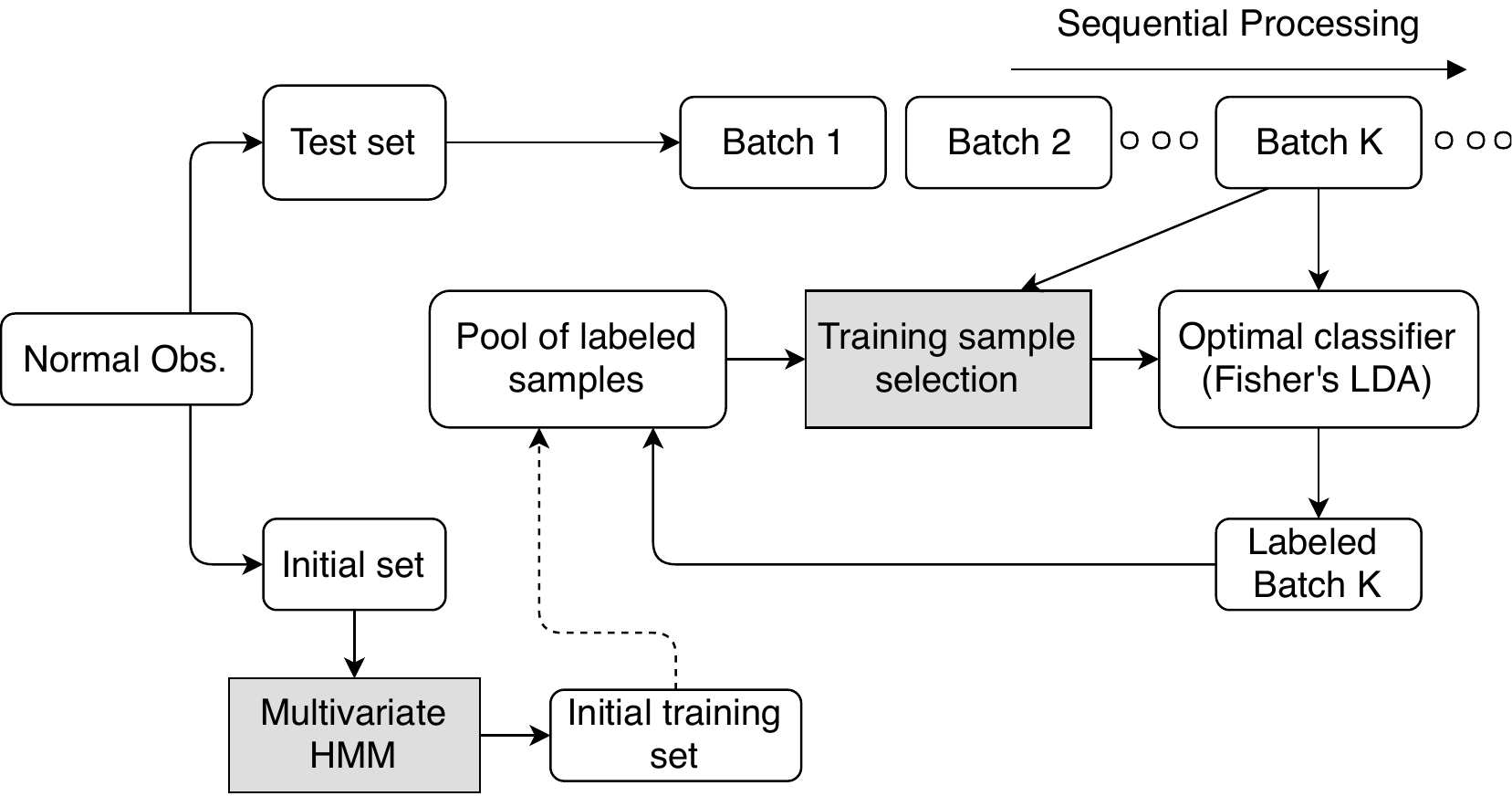}
	\label{fig:flda}}
	\subfloat[]{\includegraphics[width=.37\linewidth]{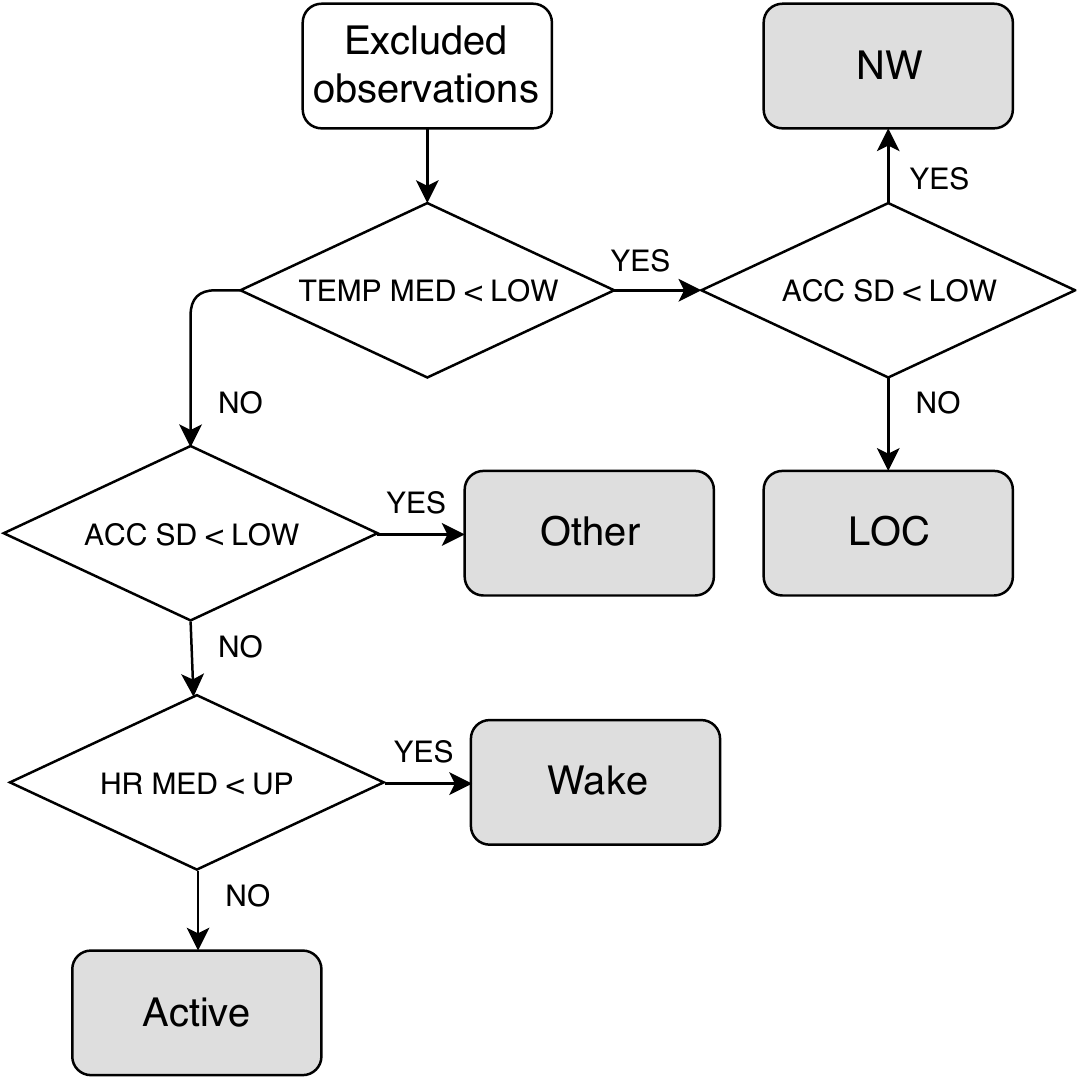}
	\label{fig:cat}}
	\caption{\textbf{(a)} a general pipeline of processing and analyzing personal wearable data; \textbf{(b)} the proposed sequential learning procedure to determine sleep/wake states; \textbf{(c)} a thresholding procedure to categorize excluded observations in the pre-processing stage of the HVC data analysis.}
	\label{fig:overview}
\end{figure*}


\IEEEPARstart{B}{eing} an essential part of the daily life cycle, sleep has repeatedly been found to be associated with immune, cardiovascular, and neuro-cognitive function \cite{miller2015role, irwin2015sleep}, among other functions. Many studies have revealed that changes in sleep pattern can be an important modulator of human response to diseases. For example, in a human viral challenge (HVC) study \cite{drake2000effects}, researchers found that nasal inoculation with rhinovirus type 23 significantly reduced total sleep time among symptomatic individuals during the initial active phase of the illness. In another study \cite{cohen2009sleep}, shorter sleep duration in the weeks preceding an exposure to a rhinovirus was found associated with lower resistance to illness. Therefore, the development of effective sleep monitoring methods has been of increasing interest.

Polysomnography (PSG) is the gold standard for sleep monitoring in sleep-related studies \cite{kaplan2017gold, dunn2018wearables}. The standard PSG procedure measures multiple bio-signals, including electroencephalogram (EEG), electrocardiogram (ECG), and electromyogram (EMG), and requires a registered technician at a clinical lab to perform sleep scoring according to the R \& K guidelines \cite{randk}. Recently much effort has been made to remedy the inconvenience of manual scoring and, alternatively, to provide automated analysis of the PSG signals, see \cite{flexerand2002automatic, malhotra2013performance, kang2018state} among others. While PSG provides relevant information about sleep sessions, it is usually excessively expensive, restrictive and cumbersome, especially when the primary goal is to monitor sleep/wake sessions outside of the lab setting. In particular, its reliance on specialized equipment only available in dedicated facilities hinders its application to field-based ambulatory studies and related clinical practice.

Thus there has been growing interest in developing cost-effective and reliable alternatives to PSG for sleep monitoring. Actigraphy (ACTG), which records movement by devices equipped with an accelerometer, is one of the most popular choices. ACTG has been widely used to study sleep/wake patterns, for example, in \cite{cole1992automatic, blood1997comparison, hedner2004novel, marino2013measuring}. More applications of ACTG to study various sleep disorders, circadian rhythm, as well as its validity in comparison to the PSG can be found in systematic review papers \cite{sadeh1995role, ancoli2003role, van2011objective}, and references therein. A critical drawback of ACTG signal analysis is that it equates sleep to immobility. Many algorithms designed for ACTG, such as in \cite{cole1992automatic} and \cite{kripke2010wrist}, use a weighted moving average of the measured movement to assign a sleep/wake state to each epoch, in which the beginning of a sleep session is typically marked by activity falling below a pre-specified threshold. In practice, however, sleep often significantly lags behind wrist immobility, leading to ACTG over-estimation of the sleep duration \cite{ancoli2003role, marino2013measuring}.

With the proliferation of multiple wearable sensors, portable multimodal systems have been developed to improve the accuracy of sleep/wake detection by integrating multiple signals besides accelerometry-measured activities.  Signals captured by wearable sensors can include blood volume pulse (BVP), heart rate (HR), electrodermal activity (EDA), temperature, respiration effort (RSP), ambient light and sound, which can be collected by wrist-worn, ankle-worn, arm-worn, lapel-worn, chest-worn and other devices such as mobile phones \cite{karlen2008improving},\cite{saeb2017scalable}, \cite{dunn2018wearables}. These methods usually require training samples with \textit{a priori} labels of true sleep/wake states, which are often provided by PSG or self-reported sleep diaries.

Existing methods of sleep/wake detection usually train a common stationary model for all subjects. This could be problematic for sleep monitoring under perturbed environments, where both inter-subject variability and intra-subject temporal variation of sleep pattern impede direct application of such methods. For example, in an HVC study, distributions of wake and sleep states may be significantly different between infected individuals and uninfected ones, and vary substantially over the whole study period due to the self-therapeutic capacity of individual immune systems. This phenomenon is known as ``data shift'' \cite{moreno2012unifying}, and methods capable of adapting to data shift belong to the transfer learning framework \cite{pan2010survey}.

In this paper, we develop an unsupervised transfer learning algorithm based on multivariate hidden Markov model (HMM) and Fisher's linear discriminant analysis (LDA) to infer the sleep/wake states of an individual under perturbed environments
using multimodal signals collected by a portable system, such as a wearable device.
The proposed algorithm does not require \textit{a priori} information about true states for training, and adaptively adjusts to sleep pattern shift by sequentially re-training models with samples most relevant to the current time point. We illustrate an application of the proposed algorithm on tracking data collected by a wearable wristband in an HVC study. Through the detected sleep/wake sessions, we discovered several physiological features strongly predictive to early-stage infection outcomes of clinical importance, including infection status and infection onset time.
Figure~\ref{fig:pipe} shows a general pipeline of processing and analyzing personal wearable data, consisting of an adaptive thresholding module for anomaly filtering and a sequential learning module (Figure~\ref{fig:flda}) based on HMM and Fisher's LDA.

The remainder of this paper is organized as follows: we introduce data pre-processing steps in Section~\ref{sec:preprocess}. Section~\ref{sec:sleep_detect} presents the proposed algorithm for sleep/wake detection in detail. Section~\ref{sec:analysis} contains an application of the proposed algorithm to wearable data collected in an HVC study, with additional results from statistical analyses of the detected sleep/wake sessions. Finally, Section~\ref{sec:discussion} contains some concluding remarks.


\section{Data pre-processing}\label{sec:preprocess}


Our algorithm aims to detect sleep/wake sessions based on multimodal physiological signals captured by a portable system, such as 3-axis acceleration, heart rate, and skin temperature.
Wearable sensors often capture data at high frequencies, and because of the fact that they are worn by subjects in non-laboratory situations, the raw data collected are often voluminous and noisy. We propose a pre-processing pipeline for the raw sensor data containing the following steps.

\begin{figure*}[ht]
  \centering
  \subfloat[]{\includegraphics[width = \linewidth, valign=t]{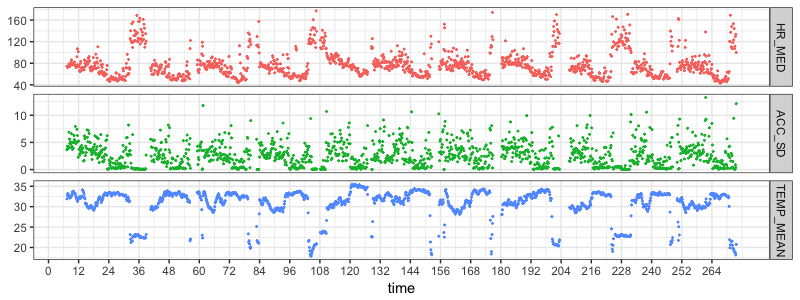}
  \label{fig:scatter_bin}}
  \hfil
  \subfloat[]{\includegraphics[width = 0.48\linewidth, valign=t]{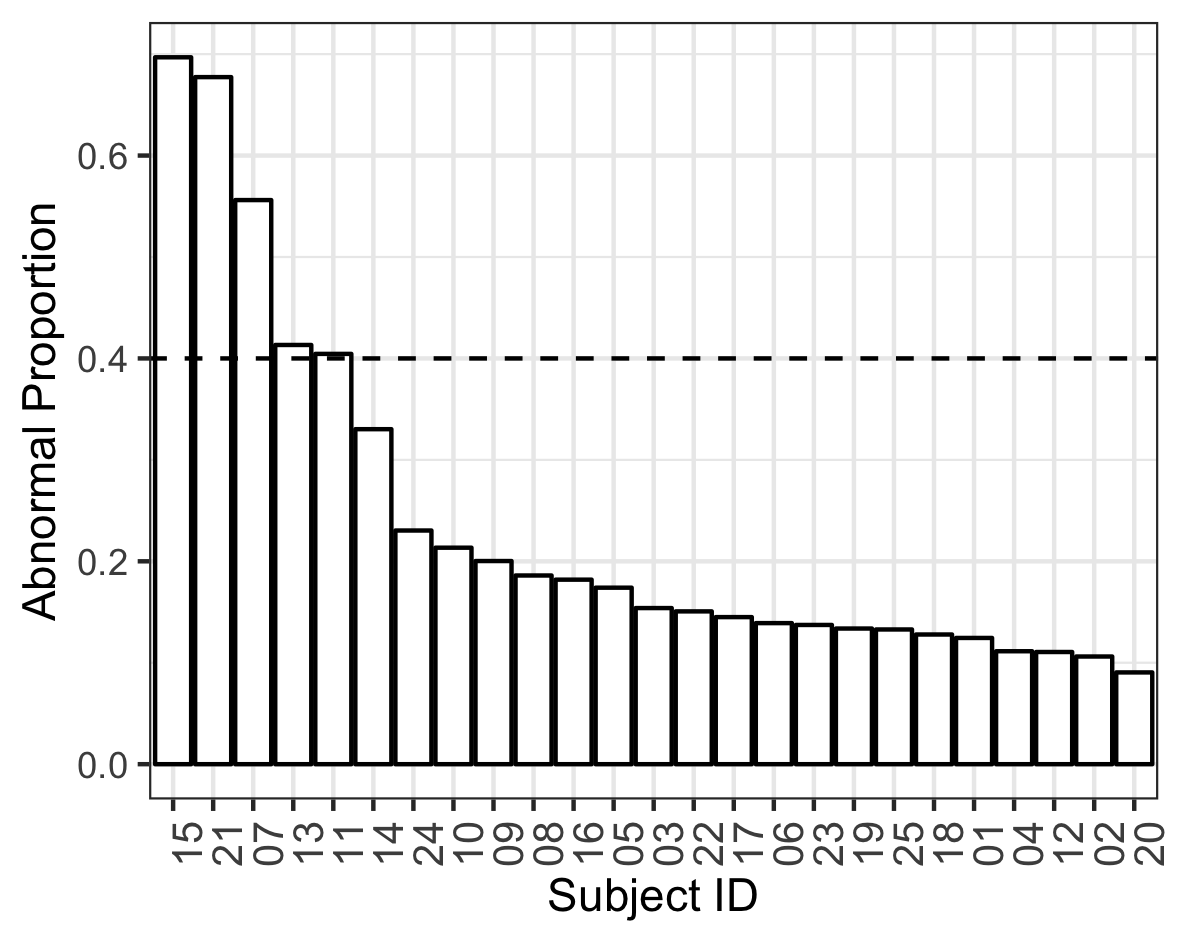}
  \label{fig:abn_hist}}
  \subfloat[]{\includegraphics[width = 0.48\linewidth, valign=t]{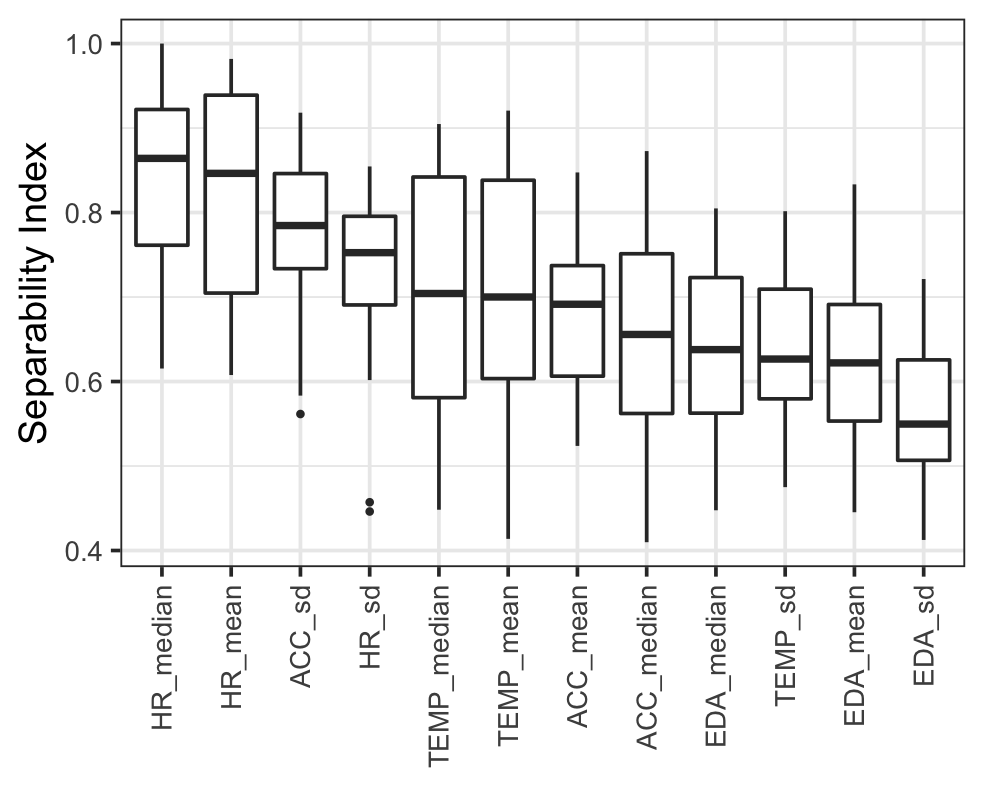}
  \label{fig:marginal_gsi}}
  \caption{(\textbf{a}) scatter plots of three summary statistics obtained from a randomly selected subject in an HVC study, including MED of heart rate (HR), SD of Euclidean norm of acceleration (ACC), and MEAN of skin temperature (TEMP); (\textbf{b}) barplot of abnormal data proportions of all 25 subjects from an HVC study; (\textbf{c}) box-plots of SI values of samples in the resting and sleeping states, using each one of the summary statistics separately.}
  \label{fig:pre-process}
\end{figure*}

\subsection{Segmentation and summarization}
To mitigate the impact of occasional poor readings and reduce computational burden, we segment the whole timeline into non-overlapping epochs of equal length. Three summary statistics, namely, mean (MEAN), median (MED) and standard deviation (SD), are extracted from each type of bio-signal within each epoch. For one epoch, if more than 90\% of the designed capacity (sampling frequency $\times$ epoch length in seconds) is not observed, such as when the device was powered-off, data from this epoch is marked as ``not available (NA)". The resulting summary statistics from the three types of signals are denoted by 
\[
  \bigl\{\bm{X}_{i,t}, \, i = 1,\dots,n, \; t = t_1, t_2, \dots, t_N\bigr\},
\]
where $t$ is the time stamp aligned to a certain origin with $t_1, t_2, \dots, t_N$ denoting respective starting times of each epoch, and $\bm{X}_{i,t}$ represents a $3 \times p$-dimensional vector of the summary statistics for subject $i$ during an epoch starting at $t$ with $p$ being the number of signals available. As an example, Figure~\ref{fig:scatter_bin} shows scatter plots of three summary statistics obtained from a randomly selected subject in an HVC study, including MED of heart rate (HR), SD of Euclidean norm of acceleration (ACC), and MEAN of skin temperature (TEMP).

\subsection{Abnormal observations filtering}\label{subsec:abn}
Summarization alone cannot handle abnormal readings that occur over an extended period of time. In the example shown in Figure~\ref{fig:scatter_bin}, it is clear that there exist abnormal values in the summarized data, such as biologically impossible low temperature (e.g. $<\, 20^\circ$C) or extremely high heart rate (e.g. $> 160$bpm). This could be due to poor contact of sensors to the subject or simply the case that the device was not worn during the time period. These abnormal readings can seriously compromise the performance of our sleep detection algorithm introduced in Section~\ref{sec:sleep_detect}, and therefore we propose an adaptive thresholding procedure using a subset of summary variables $\tilde{\bm{X}}_{i,t} = (\tilde{X}_{1i,t}, \dots, \tilde{X}_{Mi,t})$. The selected variables should have distinctive distributions under normal and abnormal situations. For example, the mean temperature is usually much lower in abnormal cases, when the sensor measures ambient temperature rather than body temperature. We then filter out data that with $\tilde{\bm{X}}_{i,t} \notin \mathcal{C}_i$, where $\mathcal{C}_i$ is a joint feasible region determined by a sequence of cutoff values $\{c_{i1}, \dots, c_{iM}\}$ for $\tilde{\bm{X}}_{i,t}$.

The subject dependent cutoff values are obtained by separately performing k-means clustering on each variable in $\tilde{\bm{X}}_{i,t}$, { and partitioning samples into three clusters, corresponding to three states: sleep, wake and abnormal. Let $\{S_{im1}, S_{im2}, S_{im3}\}$ be the resulting clusters with centers (centroids) $\{\mu_{im1}, \mu_{im2}, \mu_{im3}\}$ arranged in descending order. Consider again the example of mean temperature. The cluster $S_{im3}$ having the smallest center value was observed to consist of abnormal samples. Occasionally, the variation within abnormal samples is even larger than the difference between sleep and wake state, in which case the middle cluster may also contain abnormal samples. We identify these cases by comparing the distances between $S_{im1}, S_{im2}$ and $S_{im2}, S_{im3}$. Specifically, the set of normal, i.e., non-abnormal, samples $S_{im}^{nor}$ based on mean temperature is determined as follows:
\[
  S_{im}^{nor} = 
  \begin{cases}
    S_{im1} \cup S_{im2} & \text{if } |\mu_{im2} - \mu_{im1}| < |\mu_{im3} - \mu_{im2}|, \\
    S_{im1} & \text{o.w.}. \\
  \end{cases}
\]
With the normal sets $S_{im}^{nor}$ for $m = 1, \cdots, M$,} the cutoff values are then 
\[
  c_{im} = 
  \begin{cases}
    Q_{0.025}(\{\tilde{X}_{mi,t} \in S_{im}^{nor}\}) & \text{if } \mu_{i, nor} > \mu_{i, abn},\\
    Q_{0.975}(\{\tilde{X}_{mi,t} \in S_{im}^{nor}\}) & \text{o.w.}, \\
  \end{cases}
\]
where $Q_q(\cdot)$ calculates the $q-$quantile of samples, and $\mu_{i, nor}$, $\mu_{i, abn}$ are mean measurement values of the normal and abnormal set, respectively. To robustify our classification procedures we used a quantile-thresholding rule to exclude samples near the boundary of the normal set. We found the 2.5\% and 97.5\% quantiles to provide adequate robustness for our data but other threshold values may be selected at the discretion of the analyst. The feasible interval for $\tilde{X}_{mi,t}$ is $[c_{im}, +\infty)$ if $\mu_{i, nor} > \mu_{i, abn}$ and $(-\infty,c_{im}]$ otherwise.

\section{An adaptive sleep detection algorithm}\label{sec:sleep_detect}
We propose an adaptive sequential learning algorithm to identify sleep periods using multimodal bio-signals, which accounts for a potential sleep pattern shift due to perturbed environments. The proposed algorithm is applied to each subject separately. Thus, for simplicity of notation, we suppress the subscript $i$ in the following discussion.

\subsection{Initial training set based on hidden Markov model}
We first establish an initial training set $\mathcal{S}_{ini} = \bigl\{(\bm{x}_t,y_t),\, t = t_1, \dots, t_{n_0} \bigr\}$ using a multivariate hidden Markov model (HMM) \cite{visser2010depmixs4}, where $\bm{x}_t$ is a subset of observed summary statistics $\bm{X}_t$, relevant to the circadian pattern, and $y_t$ is the latent label indicating sleep status ($y_t = 0$ if wake; $y_t = 1$ if sleep). 
The end point $t_{n_0}$ of the initial training set is selected as the time before potential perturbation on the sleep pattern, so that we can posit a multivariate HMM as follows:
\begin{equation*}
  \bm{x}_t \mid (u_t = k) \sim N(\bm{\mu}_k, \Sigma_k), \;\;\; k = 1,\dots,K,  
\end{equation*}
where the discrete latent variable $u_t$ follows a Markov process with transition matrix $A_{K\times K}$. The variable $u_t$ is closely related to, but not necessarily the same as,  $y_t$ depending on whether $\bm{x}_t$ has multiple states during wake sessions. In fact, the sleep/wake label $y_t$ estimates are derived from the estimated latent variable $u_t$. For example, the mean heart rate may have two different states (resting and active) when the subject is awake, in which case $u_t$ has three states, i.e., $K = 3$, with both the resting and active state corresponding to $y_t = 0$. { In the following discussion, we will focus on the derived binary label $y_t$, since our primary goal is to detect sleep/wake sessions.}

To select the optimal set of $\bm{x}_t$ as well as optimal number of states $K$ for each subject, we suggest first specifying a list of possible model configurations, for example, a two-state trivariate HMM using MEAN and SD of heart rate and SD of acceleration. A series of HMMs are then fitted and evaluated by the separability index (SI) proposed in (\ref{eqn:si}) and the HMM model with the maximum SI from the pool of candidates is selected as the model for deciding sleep/wake states $y_t$ in the initial training set. 


\subsection{Sequential learning with Fisher's LDA}
Due to potentially perturbed environments, the underlying distribution of $\bm{x}_t \mid y_t$ may not be stationary over time. Thus, the initial training set cannot be directly used to determine the sleep labels for subsequent epochs. We propose performing Fisher's LDA in a sequential fashion using the same summary statistics as used in the chosen HMM, resulting in a dynamic classifier that is capable of adjusting for covariate shift. Panel \textbf{(b)} in Figure~\ref{fig:overview} demonstrates this sequential learning procedure and Algorithm \ref{alg1} below summarizes the major steps to implement it.

\begin{algorithm*}[!ht]
	\caption{Sequential learning with adaptive-size sliding window}
	\begin{algorithmic}[1]
	\State $current = t_{n_0}$  \Comment{Start from the initial training set}
	\While{$current < t_N$}    \Comment{Continue until all epochs are labeled}
		\For{$l$ in $1$ to $L$}  \Comment{Consider all window lengths}
				\Procedure{Fisher's LDA}{train = $(current - d_l,current]$, test = $(current, current + \Delta t]$}
					\State $\bm{w} = S_W^{-1}(\bar{\bm{x}}_{1\cdot} - \bar{\bm{x}}_{0\cdot})$
					\State $z_{t'} = \bm{w}^T \bm{x}_{t'}$
					\State $y_{t'} = \mathds{1}\bigl\{(z_{t'}-\bar{z}_{0\cdot})^2/\sigma^2_0 - (z_{t'}-\bar{z}_{1\cdot})^2/\sigma^2_1 > \log(\gamma \, \sigma_1^2/\sigma_0^2) \bigr\}$
					\State $\mbox{SI}(d_l) = \mbox{SI}(\mbox{train}, \mbox{test}; d_l)$   \Comment{Separability index for a given window lengths}
				\EndProcedure
		\EndFor
		\State $d_0 = \mbox{\textbf{which.max}}(\mbox{SI}(d_l))$ \Comment{Select the optimal window lengths with maximum SI}
		\State $y_{t', opt} = y_{t'} \,|\, d_0$ \Comment{Determine the labels using the optimal window}
		\State $current = current + \Delta t$  \Comment{Move on to the next test batch}
	\EndWhile
	\end{algorithmic}
	\label{alg1}
\end{algorithm*}

Fisher's LDA is a powerful tool for dimension reduction and classification \cite{friedman2001elements}. It is motivated by the idea of finding the optimal direction vector $\bm{w}$, onto which the projected scores of training samples have the largest ``separability" between classes. Let the training dataset be $\bigl\{(\bm{x}_t, y_t), \, t \in \mathcal{T}_{trn}\bigr\}$ and let $z_t = \bm{w}^T\bm{x}_t$ denote the projected scores. The separability is measured by the following ratio of \textit{between-class} variation and \textit{within-class} variation:
\begin{equation}\label{eqn:J1}
  J(\bm{w}) = \frac{(\bar{z}_{1\cdot} - \bar{z})^2 + (\bar{z}_{0\cdot} - \bar{z})^2}{\sum_{t\in G_1} (z_{t} - \bar{z}_{1\cdot})^2 + \sum_{t\in G_0} (z_{t} - \bar{z}_{0\cdot})^2} \; ,
\end{equation}
where $G_k = \{t: y_t =k, \, t \in \mathcal{T}_{tr} \}$ for $k = 0, \, 1$ denote indices of data from two classes, respectively, $\bar{z}_{k\cdot} = \sum_{t \in G_k} z_t / |G_k|$ for $k = 0, \, 1$ are the class-specific mean scores and $\bar{z}$ is the overall mean score. The optimal $\bm{w}$ that maximizes (\ref{eqn:J1}) is
\begin{equation}\label{eqn:optw}
  \bm{w} = S_W^{-1}(\bar{\bm{x}}_{1\cdot} - \bar{\bm{x}}_{0\cdot}),
\end{equation}
where $\bar{\bm{x}}_{k\cdot} = \sum_{t \in G_k} \bm{x}_t / |G_k|$ for $k = 0,\, 1$ and $S_W = \sum_{k = 0,1}\sum_{t\in G_k}(\bm{x}_t - \bar{\bm{x}}_{k\cdot})(\bm{x}_t - \bar{\bm{x}}_{k\cdot})^T$, known as the \textit{within-class} scatter matrix.
Let $\sigma^2_k = \sum_{t \in G_k}(z_t - \bar{z}_{k\cdot})^2/(|G_k|-1)$ for $k = 0, \, 1$ be the class-specific sample variances. For a test sample $\bm{x}_{t'}$, we build a classifier based on the projection vector $\bm{w}$ from Fisher's LDA and predict the label using the following rule:
\begin{equation}\label{eqn:decision_rule}
  y_{t'} = \mathds{1}\left\{\frac{(z_{t'}-\bar{z}_{0\cdot})^2}{\sigma^2_0} - \frac{(z_{t'}-\bar{z}_{1\cdot})^2}{\sigma^2_1} > \log\left(\gamma \,\frac{\sigma_1^2}{\sigma_0^2}\right) \right\},
\end{equation}
where$z_{t'} = \bm{w}^T \bm{x}_{t'}$ and $\gamma$ is a multiplicative factor for the threshold with its value set to $1$ by default in our HVC analysis. In other words, the label of a test sample depends on the relative distance of the sample to the class centroids weighted by dispersion. Assuming the conditional distribution $\bm{x}_t \mid y_t$ is multivariate normal for both classes, it is easy to show that (\ref{eqn:decision_rule}) is equivalent to the Naive Bayes classifier \cite{friedman2001elements} based on the projected score $z$.

{ While the sleep pattern may gradually shift over time, we will assume that the change is negligible within a short window. Therefore, we process and predict sleep labels incrementally within batches of non-overlapping time windows of pre-specified length $\Delta t$. This is similar to approaches used in online learning \cite{shalev2012online}, except that we use predicted labels instead of true labels, which are not available to us.}
To be more specific, after establishing the initial training set, we perform the Fisher's LDA using the same summary statistics as used in the chosen HMM and predict sleep labels for epochs coming from the next test batch $\bigl\{\bm{x}_t, \; t_{n_0} < t \leq t_{n_0} + \Delta t, \, t \in \{t_1, \cdots, t_N\}\bigr\}$.
The processed batch is then added to the pool of labeled samples so as to form an enlarged training dataset used to renew Fisher's LDA classifier. The process continues sequentially until all epochs are labeled. 

{ Inspired by methods for concept drift adaption in the supervised online learning literature \cite{widmer1996learning, klinkenberg2000detecting, gama2014survey, kuncheva2009window},}
we incorporate a taper windowing mechanism which excludes outdated training samples that are distant in time from the current batch under investigation and no longer representative. To implement this mechanism we adopt a sliding-window technique for performing Fisher's LDA in which only the most recently labeled training samples are used. The length of the window has to be carefully determined to achieve a desirable performance. { Unlike the case of supervised online learning, with no information on true labels, it is infeasible to select optimal window length by maximizing prediction accuracy in test batches. Instead, the separability index (SI) \cite{thornton1998separability, greene2001feature} is maximized as a proxy for accuracy.} Specifically, first specify a set of candidate window lengths $\bigl\{d_1, \cdots, d_L\bigr\}$. Then, letting $t_n$ denote the most recent epoch with sleep label, and creating $\mathcal{T}_{trn}(d_l) = \bigl\{t: \, t_n-d_l  < t \leq t_n,\, t \in \{t_1, \cdots, t_N\}\bigr\}$ for $l = 1,\cdots, L$ and $\mathcal{T}_{tst} = \bigl\{t: \, t_n < t \leq t_n + \Delta t, \, t \in \{t_1, \cdots, t_N\} \bigr\}$, we define the epochs in the candidate training datasets and test batch, respectively. 
For a single window of length $d_l$, perform an initial pass of Fisher's LDA with training samples $\{(\bm{x}_t,y_t),\, t \in \mathcal{T}_{trn}(d_l)\}$. Let $\hat{\bm{w}}(d_l)$ be the optimal direction vector given by (\ref{eqn:optw}). Then, predict labels with a second pass, using the resulting Fisher's LDA classifier on the test batch, and calculate the following separability index of the joint set of training and test samples: 
\begin{equation}\label{eqn:si}
  \mbox{SI}(d_l) = \frac{\sum_{t \in \mathcal{T}_{trn}(d_l) \cup \mathcal{T}_{tst}} \{y_t + \tilde{y}(t) + 1\}\mbox{ \textbf{MOD} }2}{|\mathcal{T}_{trn}(d_l)| + |\mathcal{T}_{tst}|},
\end{equation}
where $y_t$ is the sleep label for epoch starting at $t$ (predicted label if $t \in \mathcal{T}_{tst}$), and $\tilde{y}(t)$ is the (predicted) label of its nearest neighbor. In our method, nearest neighbors are determined by the `` projection distance'' defined, for samples from epoch $t$ and $t'$, as follows:
\begin{equation}\label{eqn:pdis}
  h(t,t') = |\hat{\bm{w}}(d_l)^T(\bm{x}_t - \bm{x}_{t'})|.
\end{equation}
Among the candidate window sizes, the one maximizing the criterion $\mbox{SI}(d_l)$ is used to derive Fisher's LDA that predicts the sleep labels of samples in the current test batch. 

The sleep periods are often detected as bursts, often consisting of one or more continuous sessions interrupted by short time periods. While such bursty behavior could be directly incorporated into an HMM model, e.g., a semi-Markov switching process, we take a simpler approach and smooth these short interruptions by applying a median filter on the estimated sleep state $y_t$ sequence. The whole timeline is then partitioned into consecutive continuous sleep and wake sessions. { If there exists prior information on the minimal duration of sleep/wake sessions of interest, it can be used to set a threshold for detected session length. For example, the \textit{deep sleep} period, defined as Stage 3 \& 4 of non-rapid eye movement (NREM) period, usually starts 30 minutes after sleep onset and lasts approximately 20 to 40 minutes in the first sleep cycle \cite{carskadon2005normal}. Hence, a 60-minute threshold can be applied to include only sleep sessions with at least one deep sleep period.}

\subsection{Separability index}
\begin{figure*}[ht]
  \centering
  \subfloat[$\mbox{SI}_1$ = 0.470, $\mbox{SI}_2$ = 0.605]{\includegraphics[width = 0.32\linewidth, valign=t]{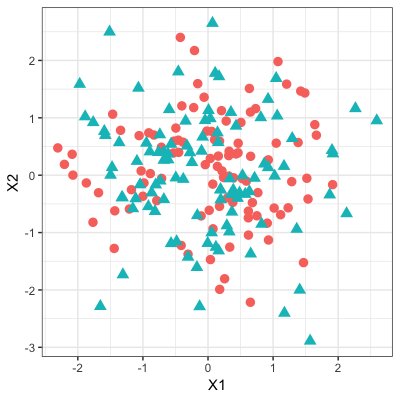}
  \label{fig:si1}}
  \subfloat[$\mbox{SI}_1$ = 0.750, $\mbox{SI}_2$ = 0.785]{\includegraphics[width = 0.32\linewidth, valign=t]{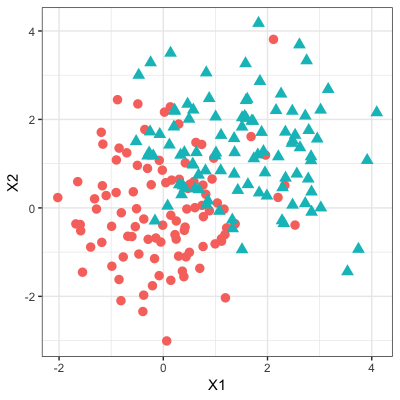}
  \label{fig:si2}}
  \subfloat[$\mbox{SI}_1$ = 1.000, $\mbox{SI}_2$ = 0.990]{\includegraphics[width = 0.32\linewidth, valign=t]{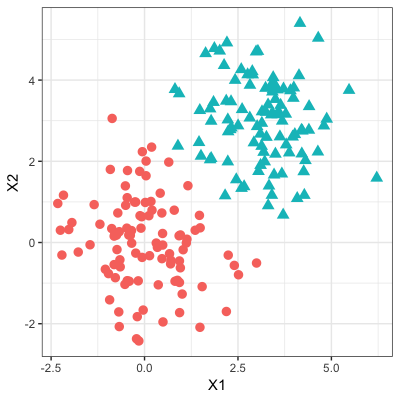}
  \label{fig:si3}}
  \caption{Separability indices under three simulated scenarios: \textbf{(a)} non-separable, \textbf{(b)} weakly separable, \textbf{(c)} strongly separable, where $\mbox{SI}_1$ is based on projection distance and $\mbox{SI}_2$ is based on Euclidean distance.}
  \label{fog:si}
\end{figure*}

By definition (\ref{eqn:si}), SI is the proportion of samples which share the same label with their nearest neighbors, and hence SI $\in [0,1]$. Intuitively, when samples from two classes form two tight, well-separated clusters with little overlap, the nearest neighbor of one sample from, say, Class 0, will most likely belong to Class 0 as well, which will result in a large SI value close to $1$. In contrast, when samples from two classes follow exactly the same distribution, i.e., completely non-separable, then the nearest neighbor of one sample will have equal probability of being Class 0 or Class 1, and thus, the SI of these samples is close to $0.5$. A large SI value implies strong separability of classes, and is usually an indication of reliable prediction. SI also depends on the measure of distance used to determine the nearest neighbors. The projection distance captures the difference among samples on the optimal direction $\bm{w}$ that is most relevant to distinguishing between the two classes, and thus is better than Euclidean distance in regard to characterizing separability, as demonstrated by Figure~\ref{fog:si}.

Figure~\ref{fog:si} illustrates three simulated cases with different levels of separability. Two variables $(X_1, X_2)$ are generated from $\text{BVN}(\mu_1, \mu_2, 1, 1, 0)$. For samples in Class 0 (indicated by red dots), $\mu_1 = \mu_2 = 0$, while for samples in Class 1 (indicated by green triangles), we considered three settings: (1) $\mu_1 = \mu_2 = 0$; (2) $\mu_1 = \mu_2 = 1.5$; (3) $\mu_1 = \mu_2 = 3$, corresponding to non-, weakly and strongly separable scenarios. SI values based on both projection distance and Euclidean distance are reported for each case. We observe that SI is indeed able to effectively characterize separability, and that the projection distance is preferred since it gives value closer to 0.5 in the non-separable case and value closer to 1 in the strongly separable case.

\section{Application to data from an HVC study}\label{sec:analysis}
We applied the proposed algorithm to an anonymized version of a dataset collected from 25 participants enrolled in a longitudinal human viral challenge (HVC) study, which was funded by DARPA under the Prometheus program. In this human challenge study,  some of the participants developed disturbed sleep patterns shortly after infection by a pathogen (H3N2). Four channels of physiological signals, including 3-axis acceleration, heart rate, skin temperature and electrodermal activity, were collected by a wrist-worn device, Empatica E4 (Empatica Inc. USA), over the time course of the study (11 days). 


\subsection{Adaptive sleep detection}
The median (MED) of heart rate and skin temperature were used in the pre-processing stage for anomaly detection. Figure~\ref{fig:abn_hist} shows the bar-plot of abnormal data proportions of all subjects, and 5 of them were excluded from further analyses due to high proportion of abnormal readings (> 40\%).

To select proper candidate summary statistics for sleep detection, we extracted data from a typical awake but resting period (21:00 - 22:00) and a sleeping period (03:00 - 04:00) for each subject. Separability indices of samples in the resting and sleeping states were then calculated marginally using one summary statistic at a time. These indices were used as measures of the difference between these states. Figure~\ref{fig:marginal_gsi} shows the box-plots of SI values for all summary statistics. We chose three summary statistics with high average SI values (> 0.7), namely, MED and SD of heart rate and SD of acceleration, as candidates for $\bm{x}_t$. These summary statistics were then used to identify sleep/wake sessions for the subjects. Note that MEAN and MED of heart rate are almost identical, so only MED is selected.

Based on the detected sleep/wake sessions, and using MED of heart rate and skin temperature and SD of acceleration, we further categorized abnormal samples excluded in the pre-processing stage into the following states:
\begin{itemize}
  \item device not worn (NW): lower-than-normal TEMP MED and lower-than-normal ACC SD;
  \item lost of contact (LOC): lower-than-normal TEMP MED and normal ACC SD;
  \item active: normal TEMP MED and higher-than-normal HR MED and ACC SD).
\end{itemize}
Here the normal ranges of these summary statistics are determined by a well established rule of thumb \cite{tukey1977exploratory} as follows:
\begin{align}
	\mbox{LOW} &= Q_{0.25} - 1.5 * (Q_{0.75}-Q_{0.25}), \nonumber\\
	\mbox{UP} &= Q_{0.75} + 1.5 * (Q_{0.75}-Q_{0.25}),
\end{align}
where $Q_{0.25}$ and $Q_{0.75}$ are the 25\% and 75\% sample quantiles of the empirical distributions over detected wake sessions. This categorization procedure is illustrated in Figure~\ref{fig:cat}.

\begin{figure*}[ht]
  \centering
  \subfloat[]{\includegraphics[width = 0.476\linewidth, valign=t]{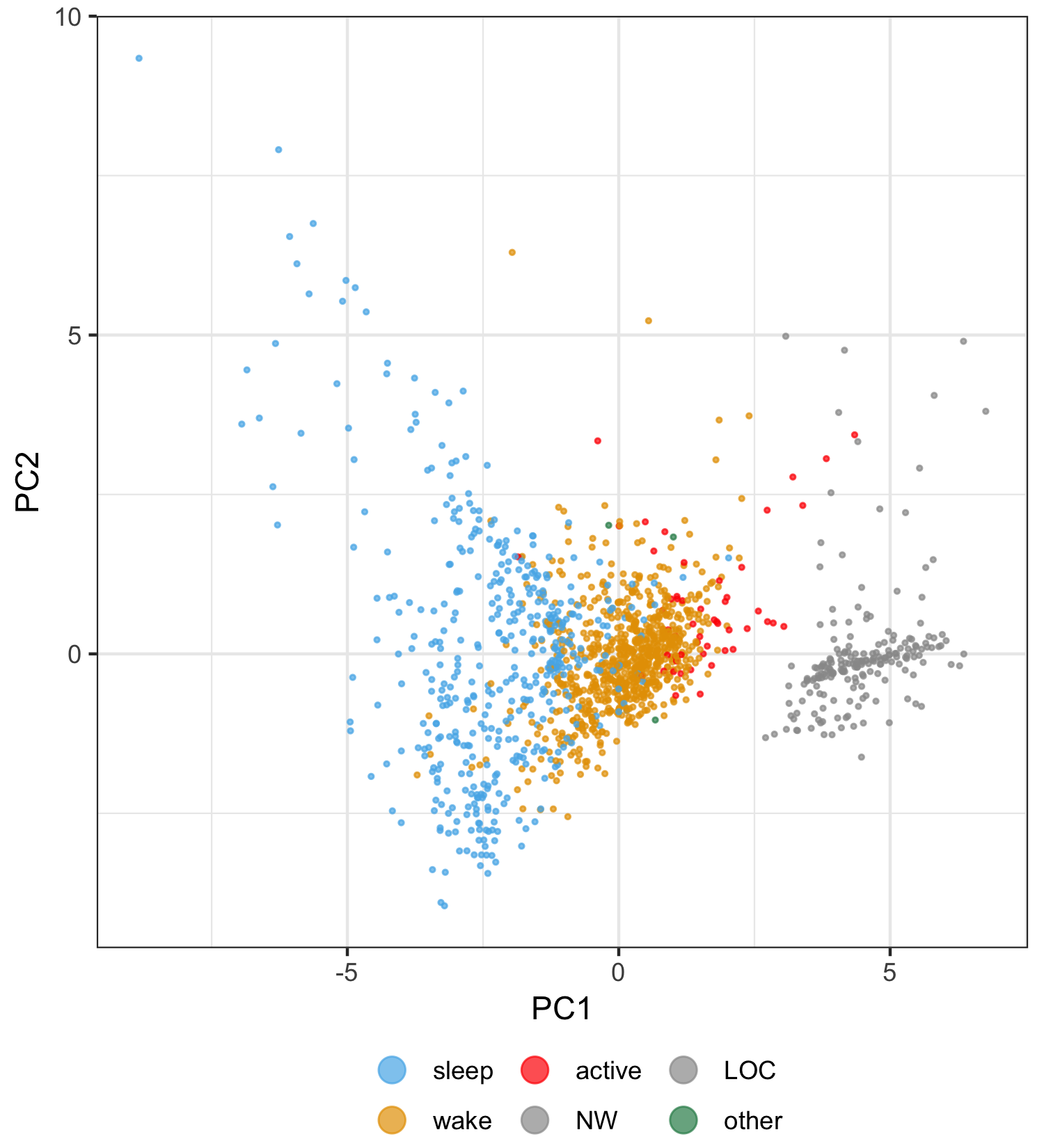}
  \label{fig:norm}}
  \subfloat[]{\includegraphics[width = 0.48\linewidth, valign=t]{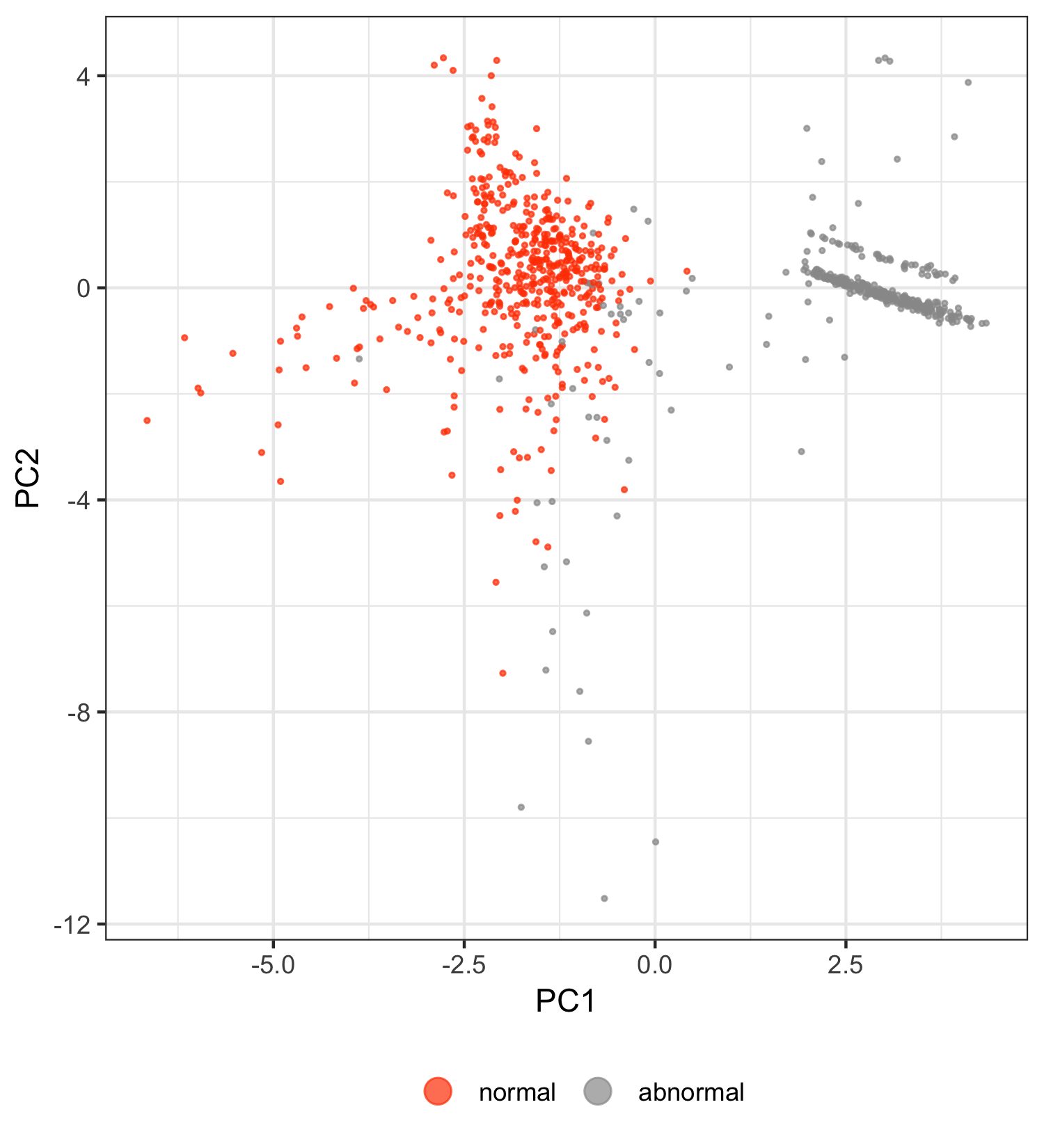}
  \label{fig:abnorm}}
  \hfil
  \subfloat[]{\includegraphics[width = 0.9\linewidth, valign=t]{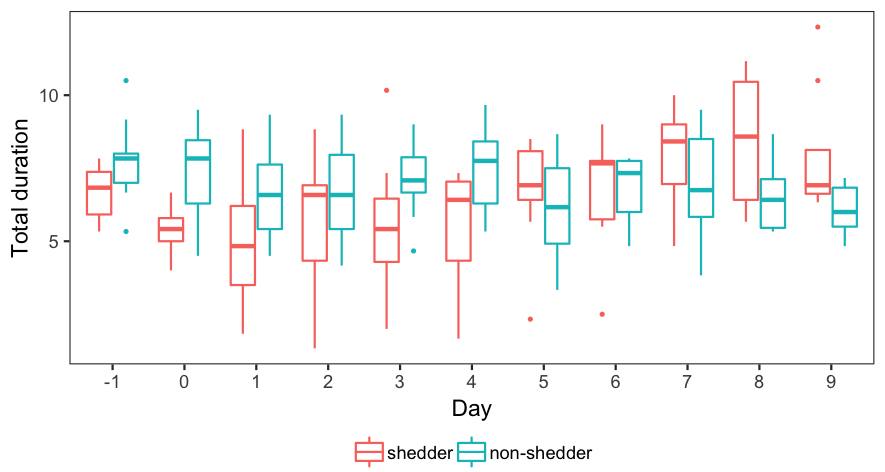}
  \label{fig:box_duration}}
  \caption{(\textbf{a}) scatter plot of the first two PCs for a subject with usable data (abnormal proportion < 40\%); (\textbf{b}) scatter plot of the first two PCs for a subject with unusable data (abnormal proportion > 40\%); (\textbf{c}) box-plots of total sleep duration for shedders and non-shedders at different points in time.}
  \label{fig:pca_subj}
\end{figure*}

We also performed principle component analysis (PCA) on the $3 \times 4$ summary statistics from all epochs for each subject separately. 
This analysis was used to confirm the effectiveness of the proposed method in detecting abnormal samples and sleep/wake sessions.
Figures~\ref{fig:norm} and \ref{fig:abnorm} show the scatter plot of the first two principle components (PC) for a subject with usable data (abnormal proportion < 40\%) and a subjects with unusable data (abnormal proportion > 40\%). For the subject with usable data, the data points form distinct clusters consistent with the categories identified by the proposed method. The subject with unusable data presents a completely different pattern in its PCA components, with the abnormal observations showing a systematic trend that corresponds to a device that is either malfunctioning, not worn properly, or not worn.

\subsection{Predictive modeling}
Finally, we considered predicting clinical outcomes with the 196 features extracted from the identified sleep/wake sessions. Each wake session is assigned to a day according to its onset time, while a sleep session is assigned to its onset day if there exists a substantial wake period (> 5 hours) prior to the session. Otherwise, it is assigned to the previous day. Table~\ref{tab:feature} summarizes the physiological features derived from sleep and wake sessions for each day, where duration and onset/offset times are exclusive features for sleep sessions, while the others are shared by sleep and wake sessions. The linear/quadratic coefficients are obtained from running linear and quadratic regression models of each summary statistic computed over a time of day. All features related to standard deviation are log-transformed.

The 20 subjects with usable data are classified according to the clinical outcome of accumulated viral shedding from the zeroth to the fifth day. Subjects in the first class are those who showed positive accumulated viral shedding, and they are called ``shedders.'' In this dichotomization, eight of 20 such subjects are shedders, while 12 others are ``non-shedders''. In addition, using a three-level ordinal scale, subjects with positive viral shedding within 48 hours after inoculation are denoted ``early shedders'' (2 out of 20); those with positive viral shedding between 48 hours and 72 hours are denoted ``mid shedders'' (6 out of 20), and others are denoted ``late shedders'' (12 out of 20, identical to non-shedders).

Figure~\ref{fig:box_duration} shows box-plots of total sleep duration for shedders and non-shedders at different points in time. We observe a clear difference between the two groups on Day 0, which is the first 24 hours after viral inoculation. The non-shedder group sleeps on the  average 2.041 hours longer than does the shedder group (p-value < 0.005) over this time period. This is consistent with studies of the effects of sleep on the course of respiratory infection \cite{drake2000effects}. The difference gradually decreases over time, and at the late stage (Day 5 and afterwards) of the study, total sleep duration of the shedders returns to the same level as that of the non-shedders.

\begin{table*}[ht]
  \centering
  \caption{Features (\textbf{196} in total) extracted from sleep (\textbf{100}) and wake (\textbf{96}) sessions for each day from 20 subjects in an HVC study.}
  \label{tab:feature}
  \resizebox{.95\textwidth}{!}{%
  \begin{tabular}{l l l}
    \hline\hline
    Name  & Number  & Description \Tstrut\Bstrut\\[+3pt]
    \hline
    Duration  & 2   & total sleep, night sleep \Tstrut\\[+3pt]
    Onset/offset  & 2   & night sleep only \\[+3pt]
    HR summary  & 9$\times$2   & 3 (mean, median, s.d.) $\times$ 3 (mean, median, s.d. within session) $\times$ 2 (sleep, wake)\\[+3pt]
    HR linear coef.   & 6$\times$2 & 3 (mean, median, s.d.) $\times$ 2 (coef.0, coef.1) $\times$ 2 (sleep, wake)\\[+3pt]
    HR quadratic coef.   & 9$\times$2 & 3 (mean, median, s.d.) $\times$ 3 (coef.0, coef.1, coef.2) $\times$ 2 (sleep, wake)\\[+3pt]
    TEMP  & 24$\times$2   & same as HR\\[+3pt]
    ACC  & 24$\times$2   & same as HR\\[+3pt]
    EDA  & 24$\times$2   & same as HR\\[+3pt]
    \hline\hline
  \end{tabular}%
  }
\end{table*}

To further assess the predictive power of these extracted features, we adopt two regression models. The first regression model is used to predict whether or not a subject sheds (binary label) and the second model is used to predict a subject's shedding onset time (ordinal label).
For the binary case, we separately fit marginal \textit{logistic} regression model 
on each feature, which models the marginal probability of shedding
\[
  \mbox{logit}\bigl[Pr(Y = 1)\bigr] = \beta_0 + \beta_1X \,,
\]
where $\mbox{logit}(x) = \log[x/(1-x)]$, $y\in \{0,1\}$ is the shedding status, and $X$ is one of the 196 features. For the ordinal onset time case, we fit a marginal \textit{continuation-ratio} model \cite{agresti2010analysis} ($y = 1$ for early shedders, $y=2$ for mid shedders and $y=3$ for late shedders), which models the conditional probabilities:
\[
  \mbox{logit}\bigl[Pr(Y = j \,|\, Y \geq j)\bigr] = \beta_0 + \beta_1X \,,
\]
for $j = 1, 2$. This may be interpreted as a discrete version of the Cox regression model. For both the logistic and the continuation-ratio models, the predictive capability of individual features is evaluated by \textit{leave-one-out cross-validation} (LOOCV). Top predictive features are reported in Table~\ref{tab:top_features}, together with their corresponding regression coefficient estimates and the Area Under Curve (AUC) based on LOOCV. For the continuation-ratio, since there exist three classes, the AUCs are calculated using the ``one-versus-the-rest'' strategy \cite{bishop2006pattern}.

\begin{table*}[ht]
  \centering
  \caption{Top physiological features capable of predicting class  membership within 24 hours of inoculation.}
  \label{tab:top_features}
  \resizebox{.95\textwidth}{!}{%
  \begin{tabular}{l c c c l c c c c}
    \hline\hline
    \multicolumn{3}{c}{\textit{logistic regression model}} &   & \multicolumn{5}{c}{\textit{continuation-ratio regression model}}\Tstrut\\[+3pt]
    \cline{1-3}\cline{5-9}
    \multirow{2}{*}{Feature}   & \multirow{2}{*}{Coef.}   & \multirow{2}{*}{AUC}   & \multirow{2}{*}{ }   & \multirow{2}{*}{Feature}   & \multirow{2}{*}{Coef.}   & \multicolumn{3}{c}{AUC}\Tstrut\\[+3pt]
    \cline{7-9}
    & & & & & & Early & Mid & Late \Tstrut\\[+3pt]
    \cline{1-3}\cline{5-9}
    HR MED.sd (sleep) & -6.196 & 0.833 & & HR MED.sd (sleep) & -7.195 & 0.944 & 0.631 & 0.844\Tstrut\\[+3pt]
    Offset & -1.629 & 0.802 & & Total duration & -1.139 &	0.750 &	0.619 &	0.813\\[+3pt]
    Total duration & -1.211	& 0.781 & & Offset & -0.864	& 0.611	& 0.488	& 0.802\\[+3pt]
    HR MEAN.sd (sleep) & -4.958 & 0.750 & & HR MEAN.sd (sleep) & -4.597	& 0.667	& 0.595	& 0.750\\[+3pt]
    Night duration & -0.896 &	0.688 & & Night duration & -0.854	& 0.667	& 0.560	& 0.698\\[+3pt]
    \hline\hline
  \end{tabular}%
  }
\end{table*}

The logistic and continuation-ratio models give consistent results that identify similar features as being important for prediction. Interestingly, all the top features come from sleep sessions. This may be due to the fact that data collected during sleep periods are less variable, due the absence of external factors such as personal behavior and changing environment, and thus better reflect underlying physiological processes. Features related to sleep heart rate variation (HR MED.sd, HR MEAN.sd) and sleep duration (total duration, offset, night duration) are the most predictive. It is remarkable that using a single non-invasive marker from the wearable device leads to AUC over $.80$ in both logistic and continuation-ratio models.
Table~\ref{tab:cor_top} shows the Pearson correlation coefficients between pairs of the top three features (total duration, offset and HR MED.sd). The total sleep duration and sleep offset time have a strong positive correlation, and they are both somewhat correlated with HR MED.sd. When all three features are included, the logistic model achieves $\mbox{AUC} = 0.844$, and the continuation-ratio model achieves $\mbox{AUC} = 0.885$ for predicting ``Late shedder versus others''. Combining the three features together leads to a small improvement in prediction performance, which is possibly due to correlation between these three features or limited sample size.
\begin{table}[ht]
  \centering
  \caption{Pairwise Pearson correlations among top three features.}
  \label{tab:cor_top}
  \begin{tabular}{ l  c c c}
    \hline\hline
                   & Total duration & Offset & HR MED.sd \Tstrut\Bstrut\\[+3pt]
    \hline
    Total duration & 1               & 0.778  & 0.434 \Tstrut\\[+3pt]
    Offset         &                 & 1      & 0.485 \\[+3pt]
    HR MED.sd      &                 &        & 1 \\[+3pt]
    \hline\hline
  \end{tabular}
\end{table}

\section{Discussion and Conclusions}\label{sec:discussion}
We have developed an adaptive personalized algorithm for detection and characterization of sleep patterns based on combining multimodal physiological signals with data covariate shift. The proposed algorithm operates without requiring \textit{a priori} information about true sleep/wake states and is capable of automatically detecting anomalies and abnormal data records. We applied the proposed algorithm to analyze physiological signals collected in an HVC study. The algorithm effectively adjusted for the inter-subject and intra-subject temporal variation during the study period, and extracted strong and non-invasive features to predict infection outcomes.

The proposed algorithm is not without limitations. By the nature of the data covariate shift adaption procedure, we must assume that the extent of shift is appropriately limited. The authors of \cite{helmbold1994tracking} mathematically characterized the extent of shift and showed that an algorithm can only adapt to data covariate shift if the rate of shift is smaller than a certain upper bound. For example, when the infection induces an abrupt increase in sleep heart rate of a subject, the classifier built on data from previous day may fail to detect a sleep session. A possible remedy would be to exclude those periods with no wake or sleep state detected, and to process data from such periods separately by applying the same hidden Markov model used in segmenting the initial training set. 

Another limitation of our study is that the HVC data used to illustrate the proposed algorithms has no ground truth information about the true sleep/wake states of the subjects. A controlled experiment performed in a sleep lab could provide such ground truth for a small number of subjects. It would be worthwhile to test the algorithm in a large scale experiment that collects self-reported sleep diaries in addition to clinical data. While sleep diaries are not 100\% reliable, such an experiment would allow us to further validate the value of the proposed unsupervised learning algorithm.
The idea of adaptive sequential learning proposed here can be easily generalized to other practical problems with data shift, beyond the current setting of sleep monitoring. With the emergence of wearables and other digital health technologies, the continued development of such integrated algorithms for continuous tracking data will lead to more powerful behavioral assessment tools, and in a broader sense, advance personalized health care.

\section{Disclosures}

\noindent{\bf Human Subjects}:
The data analysis reported in Sec. IV of this paper followed a protocol that was submitted to the Internal Review Boards of Duke University and the University of Michigan. On 4/25/2017 the Internal Review Board of Duke University Health Sciences (DUHS IRB) determined that the protocol meets the definition of research not involving human subjects, as described in 45CFR46.102(f), 21 CFR56.102(e) and 21CFR812.3(p), and that the protocol satisfies the Privacy Rule, as described in 45CFR164.514.  On 8/4/2017 the Internal Review Board of the University of Michigan (UM IRB) determined that the research performed under this protocol was not regulated.  

\vspace{0.1in}
\noindent{\bf Conflict-of-interest}:
The authors have no conflicts of interest to disclose.

\ifCLASSOPTIONcaptionsoff
  \newpage
\fi



\bibliographystyle{IEEEtran}
\bibliography{IEEEabrv,myref}
\end{document}